\documentclass{elsarticle}
\usepackage{graphicx}
\usepackage{amsmath}
\usepackage{amssymb}
\usepackage{CJK}
\usepackage[utf8]{inputenc}
\usepackage{bm}
\usepackage{MnSymbol} 
\usepackage[multiple]{footmisc}
\usepackage{tikz}
\usetikzlibrary{cd}
\usepackage{hyperref}

\graphicspath{{./Figures/}}

\def\be#1\ee{\begin{align}\begin{split}#1\end{split}\end{align}}
\def\beq#1\eeq{\begin{align}\begin{split}#1\end{split}\end{align}}

\def\CN{{\cal N}}

\def\SU{\mathrm{SU}}
\def\U{\mathrm{U}}

\newcommand{\q}{{\mathsf q}}
\newcommand{\p}{{\mathsf p}}

\begin{document}

\begin{CJK*}{UTF8}{min}

\begin{frontmatter}

\title{Integrability As Duality: \\ \medskip The Gauge/YBE Correspondence}

\author{Masahito Yamazaki (山崎雅人)}
\cortext[mycorrespondingauthor]{Corresponding author}
\ead{masahito.yamazaki@ipmu.jp}
\address{Kavli IPMU (WPI), UTIAS, The University of Tokyo, Kashiwa, Chiba 277-8583, Japan}


\begin{abstract}
The Gauge/YBE correspondence states a surprising connection between solutions to the Yang-Baxter equation with spectral parameters and partition functions of supersymmetric quiver gauge theories.
This correspondence has lead to systematic discoveries of new integrable models based on quantum-field-theory methods. We provide pedagogical introduction to the subject and summarizes many recent developments. This is a write-up of the lecture at the String-Math 2018 conference.
\end{abstract}

\end{frontmatter}

\tableofcontents

\section{Introduction} \label{sec.intro}

In this review I would like to summarize some exciting recent developments in the area
of applying gauge-theory ideas, and in particular Seiberg dualities, to integrable models.\footnote{For presentation slides, see \newline \url{http://www.tfc.tohoku.ac.jp/wp-content/uploads/2018/06/05_Yamazaki_2018SRM-E02.pdf}.} 


In the literature there are several different (although related) characterizations of integrable models.
In this lecture integrable models are defined to be the solutions of
the Yang-Baxter equation \cite{Yang:1967bm,Baxter:1972hz} (YBE) {\it with spectral parameters}.

\begin{figure}[htbp]
\centering{\includegraphics[scale=0.18]{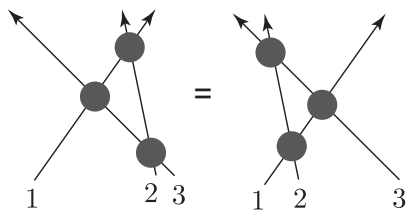}}
\caption{Graphical representation of the YBE.}
\label{Fig1}
\end{figure}

\begin{figure}[htbp]
\centering{\includegraphics[scale=0.18]{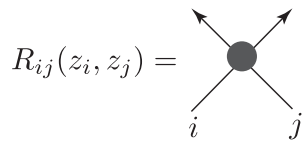}}
\caption{We associate an R-matrix at the crossing of two rapidity lines.}
\label{Fig1_2}
\end{figure}

The graphical representation of YBE is given in Figure \ref{Fig1}.
Here we have three intersecting lines (often called rapidity lines) intersecting. We label them by $1,2,3$,
and we associate three vector spaces $V_1, V_2, V_3$ for each line.
At each crossing of the two lines (say $i$ and $j$)
we associate the so-called R-matrix (see Figure \ref{Fig1_2}):
\be
R_{ij} \in \textrm{End}(V_i\otimes V_j) \;.
\ee

The YBE states that the product of three R-matrices, where the product is taken in two different orders,
coincide:
\be
&R_{23}R_{13} R_{12} = R_{12} R_{13} R_{23}  \in \textrm{End}(V_1\otimes V_2\otimes V_3) \;.
\ee
Here for example $R_{12}\in \textrm{End}(V_1\otimes V_2)$ is extended to a operator
$R_{12} \otimes 1_{V_3} \in \textrm{End}(V_1\otimes V_2\otimes V_3)$, which for simplicity we denoted by the same symbol $R_{12}$.

In this lecture we are interested in YBE with spectral parameters. This means that we have three parameters $z_1, z_2, z_3$
for the three lines $1,2,3$, and the R-matrix at the intersection of rapidity lines $i$ and $j$ depends on the associated spectral parameters $z_i$ and $z_j$:
\be
R_{ij}=R_{ij}(z_i, z_j)  \in \textrm{End}(V_i\otimes V_j) \;.
\ee

In almost all of the known integrable models,
we have the property that the R-matrix depends only on the differences between the spectral parameters: 
\be
R_{ij}(z_i, z_j)=R_{ij}(z_i-z_j) \;.
\label{rapidity_difference}
\ee
This is known as the rapidity difference property of the R-matrix.
This property holds for almost all the models discussed in this lecture,
and in the following we will assume this property unless stated otherwise.

The YBE, now with spectral parameters, reads
\be
&R_{23}(z_2-z_3) R_{13}(z_1-z_3) R_{12}(z_1-z_2)  \\
& \quad = R_{12}(z_1-z_2)  R_{13}(z_1-z_3) R_{23}(z_2-z_3)  \in \textrm{End}(V_1\otimes V_2\otimes V_3) \;.
\label{YBE2}
\ee

Once we have the YBE, then the textbook-construction of the integrable models (see e.g.\ \cite{Baxter:1982zz}) ensures
that transfer matrices commute with each other, and hence by expanding with respect to the spectral parameters $z$ 
we learn that the model has infinitely-many commuting conserved charges.
This is another definition of integrability. For this reason YBE has been studied rather intensively in integrable model literature.

While there are many important questions in the field of integrable models,
one of the most fundamental questions is the following: 
\begin{equation*}
\textit{Why integrable models exist?}
\end{equation*}

This is a highly non-trivial question. For example, if one naively tries to solve the YBE \eqref{YBE2}
one finds that the equations are over-constrained in general. To see this, let us consider the situation
where all the vector spaces $V_i$ are $N$-dimensional. Then the R-matrix, associated with the crossing of two lines,
has four boundary lines and hence has $O(N^4)$ components. By contrast, the YBE, associated with a figure with six 
boundary lines, has $O(N^6)$ components. This over-counting problem is especially severe when $N$ is large, say when $N$ is infinity
(namely when we consider an infinite-dimensional representation),
which is indeed the case in many of the models discussed in this lecture.

It is therefore a fundamental question to understand why integrable models can exist at all.
Once we have a good understanding of this one can then try to understand the pattern of existing integrable models
in the literature, and moreover try to find new integrable models based on such understanding.

I myself has been fascinated by this question over the past years, and have worked mainly on two different approaches, both based on quantum-field-theory ideas.

One approach is based on the ``4d Chern-Simons theory''. This approach was pursued first by Costello 
in 2013 \cite{Costello:2013zra,Costello:2013sla}, and more recently developed further in by Costello, Witten and myself \cite{Witten:2016spx,Costello:2017dso,Costello:2018gyb,Costello:2019tri,PartIV}.

Another approach is based on supersymmetric quiver gauge theories.
This approach is called {\it the Gauge/YBE correspondence}, since this correspondence claims rather direct relation between the YBE and the partition functions of gauge theories. This approach 
was initiated in 2012-2013 in my papers (partly in collaboration with Terashima) \cite{Yamazaki:2012cp,Terashima:2012cx,Yamazaki:2013nra}.
Our work is based on several previous important ideas in the literature, most notably the integrable models (known as the ``master solution'') 
constructed by Bazhanov and Sergeev \cite{Bazhanov:2010kz,Bazhanov:2011mz}.\footnote{There are by now substantial literature on this subject, see e.g.\ \cite{Spiridonov:2010em,Xie:2012mr,Yagi:2015lha,Yamazaki:2015voa,Gahramanov:2015cva,Kels:2015bda,Gahramanov:2016ilb,Bazhanov:2016ajm,Maruyoshi:2016caf,Spiridonov:2016uae,Yamazaki:2016wnu,Gahramanov:2016sen,Yagi:2017hmj,Kels:2017toi,Jafarzade:2017fsc,Gahramanov:2017idz,Gahramanov:2017ysd}.}
By now I have succeeded in reproducing their construction from scratch, and generalized the master solution in a number of different directions,
based on quantum-field-theory techniques.

In the rest of this lecture I will discuss the second approach to this subject.

\section{Basic Ingredients}

While there are many technical complications needed in the actual discussions of the subject,
the fundamental ideas behind the Gauge/YBE correspondence are rather simple,
and can be listed as follows:
\begin{itemize}
\item a statistical lattice as a quiver diagram
\item supersymmetric localization
\item Seiberg(-like) duality
\end{itemize}

\subsection{Statistical Lattice as Quiver Diagram}

\subsubsection{Statistical Mechanical Model}

\begin{figure}[htbp]
\centering{\includegraphics[scale=0.4]{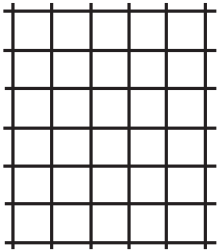}}
\caption{Statistical lattice where a statistical-mechanical model is defined. We can also regard this as a quiver diagram, a defining data for the quiver gauge theory.}
\label{Fig2_1}
\end{figure}

In textbook presentation of statistical mechanics, one starts with a statistical lattice,
such as the one shown in Figure \ref{Fig2_1}. Let us denote the set of vertices by $V$,
and edges by $E$.

In the statistical-mechanical model, 
the dynamical degrees of freedom of the theory,
namely `spins', are placed at the vertices of the lattice: $s_v$ at vertex $v\in V$.
The spins here can take values in any set and can be discrete or continuous.
For example $s_v=\pm 1$ for the Ising model. In this lecture we will encounter 
more complicated examples where $s_v$ takes values in e.g.\ $(\mathbb{R}\times \mathbb{Z}_r)^{N-1}$
for integers $r, N$ (with $N>1$) \cite{Yamazaki:2013nra}: in this example we $s_v$ has multiplet components,
and has both discrete and continuous components.

The partition function of the statistical-mechanical model is defined as
\begin{align}
Z:= \sumint \left( \prod_{v \in V} ds_v \right) e^{-\mathcal{E}\left( \{ s_v \}_{v\in V}\right)}  \;.
\label{Z_stat}
\end{align}
Here the symbol $\sumint \prod_{v \in V} ds_v$
denotes the summation over the spin configuration $\{s_v \}_{v\in V}$.
When $s_v$ has only discrete (continuous) components this should be understood as
$\sum_{\{s_v \}_{v\in V}}$ ($\int \prod_{v \in V} ds_v$), and in general we sum (integrate)
over the discrete (continuous) part of the spins.

Inside the integrand of \eqref{Z_stat} we have the so-called Boltzmann weight
$\mathcal{E}\left( \{ s_v \}\right)$, a function of spins
given as a sum of contributions from edges and vertices:
\be
\mathcal{E}\left( \{ s_v \}_{v\in V} \right)=\sum_{v \in V} \mathcal{E}^v (s_v)
+\sum_{e \in E} \mathcal{E}^e\left( \{ s_v \}_{v\in e} \right) \;,
\label{L_stat}
\ee
where $v\in e$ means that the vertex $v$ is one of the two endpoints of the edge $e$.
The function $\mathcal{E}^e$ specifies the nearest-neighbor interaction of the spins,
while $\mathcal{E}^v (s_v)$ specifies the local interaction at the vertex (such as the magnetic field dependence in the Ising model,
or the self-interaction between different components of the spin).

\subsubsection{Quiver Gauge Theory}\label{sec.2.1}

Instead of starting with a statistical-mechanical model, we would like to start with a
quantum field theory defined from the statistical lattice.

In high energy physics there are several ideas along these lines.
For example, the lattice gauge theory indeed is defined on a discrete lattice.
We here instead discuss the concept of quiver gauge theory.\footnote{The discussion of the quiver gauge theory here is 
somewhat more general than the more standard usage of the word---for example, the matter fields are not necessarily in the bifundamental or adjoint representation.
We will eventually specialize to the more standard construction, see section \ref{sec.details}.}

In quiver gauge theory we regard the statistical lattice
as the ``quiver diagram'', which specifies the gauge/matter content of the theory.
Namely, 
\begin{itemize}

\item For each vertex $v\in V$ we associate a gauge field $A_v$ valued in the gauge group $G_v$.
This means that the total gauge group of the theory is 
\be
G_{\rm total}=\prod_{v\in V} G_v \;.
\ee
\item For each edge $e \in E$ with two endpoints given by $v, v'\in V$
we associate a matter field $\phi_e$ which transforms non-trivially under 
some representation of $G_v\times G_{v'}$, but otherwise
trivially under $G_{v''}$ for $v''\ne v, v'$. 

\end{itemize}

In this discussion, the statistical lattice is simply a mnemonic for the 
gauge/matter content of the theory,
and does not have any direct relation with the actual spacetime
where the quantum field theory is defined.\footnote{For this reason it is sometimes said that the 
quiver diagram lives in the ``theory space''.}

The partition function of this quantum field theory is defined by the path-integral:
\be
Z=\int \prod_v \mathcal{D} A_v \prod_e \mathcal{D} \phi_e \,\, e^{-L\left(\{A_v \}_{v\in V},  \{\phi_e \}_{e\in E}\right)}  \;,
\label{Z_QFT}
\ee
where $\mathcal{D} A_v$ and $\mathcal{D} \phi_e$ denote the path-integral measures 
for the quantum fields $A_v$ and $\phi_e$.

The integrand, namely the Lagrangian $L$, is a functional of the 
field configurations $\{A_v \}_{v\in V},  \{\phi_e \}_{e\in E}$, and is given by
\be
L\left(\{A_v \}_{v\in V},  \{\phi_e \}_{e\in E}\right)
=\sum_{v\in V} L^v\left(A_v \right)
+
\sum_{e\in E} L^e\left(\{A_v \}_{v\in e}, \phi_e \right) \;.
\label{L_QFT}
\ee
Here the term $L^v$ is the kinetic term for the gauge field $A_v$,
and the term $L^e$ is the Lagrangian for the field $\phi_e$,
which couples non-trivially to the gauge fields $\{A_v \}_{v\in e}$.

\subsubsection{Comparison}

There is a striking parallel between the two stories above, between statistical-mechanical models
and quiver gauge theories. For example, the role of the spins $s_v$ are played by the gauge fields
$A_v$, and in the partition functions \eqref{Z_stat} and \eqref{L_stat} are replaced by \eqref{Z_QFT} and \eqref{L_QFT}, respectively.

Encouraged by this parallel, one can try to construct the statistical-mechanical model from quantum field theory.

There are still important differences between the two stories, however.

\begin{itemize}
\item In statistical mechanics all the degrees of freedom (namely spins $s_v$) live at the vertices.
Quiver gauge theory, by contrast, has matter fields $\phi_e$ located at the edges, in addition to the gauge fields $A_v$ located at the vertices. In other words, we have a non-standard 
statistical-mechanical model where we have degrees of freedom both at the edges and the vertices.

\item In quiver gauge theory the partition function \eqref{Z_QFT} is defined by the path integral,
and does not necessarily fit with the more conventional discussion of statistical-mechanical models \eqref{Z_stat},
where the partition function is defined by a finite integral/sum.

\end{itemize}

In the next subsection we shall explain how these differences are removed with the help 
of supersymmetry.

\subsection{Supersymmetric Localization}\label{sec.2.2}

The second ingredient of the Gauge/YBE correspondence is supersymmetric localization.

Suppose that we supersymmetrize our quiver gauge theory.
The details of how this works depends of course on the 
dimensionality and the number of supersymmetries.

While the story below will work in general, let us 
for concreteness consider the case of four-dimensional
quiver gauge theory with $\CN=1$ supersymmetry (namely four supercharges).
Then we have

\begin{itemize}

\item For each vertex $v\in V$ we associate an $\CN=1$ vector multiplet $V_v=(A_v, \lambda_v, F_v)$,
whose on-shell degrees of freedom contains a gauge field $A_v$, a gaugino $\lambda_v$, and an auxiliary field $F_v$,
all in the adjoint representation of the 
gauge group $G_v$.

\item For each edge $e \in E$ with two endpoints given by $v, v'\in V$
we associate an $\CN=1$ chiral multiplet $\Phi_e=(\phi_e, \psi_e, D_e)$ in a non-trivial representation of $G_v\times G_{v'}$:
this multiplet contains a complex scalar $\phi_e$, a fermion $\psi_e$, and an auxiliary field $D_e$. 

\end{itemize}

The definition of the partition function is of course then supersymmetrized:
\be
Z& =\int \prod_v \mathcal{D} V_v \prod_e \mathcal{D} \Phi_e  \,\, e^{-\mathcal{L}\left(\{V_v \}_{v\in V},  \{\Phi_e \}_{e\in E}\right)}  \;.
\label{Z_SUSY}
\ee
where we denoted the supersymmetrized path-integral measure as
\be
\mathcal{D} V_v = \mathcal{D} A_v \mathcal{D} \lambda_v  \mathcal{D} F_v \;, \quad
 \mathcal{D} \Phi_e = \mathcal{D} \phi_e  \mathcal{D} \psi_e \mathcal{D} D_e \;.
\ee

Then partition function \eqref{Z_SUSY}, when considered in the flat spacetime, is in general a divergent quantity.
We can however regularize the IR divergence by placing the theory on a certain compact manifold $M$,
and after suitably regularizing the UV divergence we can 
define a well-defined partition function $Z[M]$;\footnote{We need to keep track of the counterterms to see if $Z[M]$ are indeed regularization independent.
Typically supersymmetric further constrains the possible counterterms, thus making the partition function well-defined.} we use the same formula \eqref{Z_SUSY} but now everything is defined on $M$.

Now the supersymmetric partition function states that we can compute the partition function $Z[M]$,
which is a priori defined as an infinite-dimensional path-integral, reduces to 
a finite-dimensional integral/sum:
\be
Z& = \sumint \prod_{v\in V} d\sigma_v  \,\, e^{-\mathcal{L}(\sigma_v)} \;.
\label{Z_localize}
\ee

In \eqref{Z_localize} $\sigma_v$ is a finite-dimensional variable associated at the vertex $v\in V$ (namely associated with the gauge group $G_v$),
and as we will see can be continuous or both continuous and discrete variables depending on the choice of the manifold $M$. In the examples below
$\sigma_v$ arises from the holonomies of the gauge field along the non-trivial cycles of $M$.
Note that there is no integral/sum variable associated with the edges of the graph, i.e.\ the matter fields $\phi_e$ vanish at the saddle point locus. 
While this is not guaranteed in general supersymmetric localization,
this is indeed the case in all the examples we study in this lecture, and in any case in the following we will assume this.\footnote{In supersymmetric localization the
path-integral reduces to an integral/sum over the saddle point configurations. If all the scalars in the matter multiplets $\phi_e$ are trivial on this saddle point then
there is edge variables in \eqref{Z_localize}.}

Finally, the integrand is $\mathcal{L}(\sigma_v)$ takes the form
\be
\mathcal{L}\left(\{\sigma_v \}_{v\in V} \right)
=\sum_{v\in V} \mathcal{L}^v\left( \sigma_v  \right)
+
\sum_{e\in E} \mathcal{L}^e\left(\{\sigma_v \}_{v\in e} \right) \;.
\label{L_localize}
\ee
Here $\mathcal{L}^v$ ($\mathcal{L}^e$) denotes the 
classical plus one-loop contributions from the 
gauge fields (matter fields). This formula should be compared with the Boltzmann weight for the 
statistical-mechanical model \eqref{L_stat}.

\subsubsection{Some More Details}\label{sec.details}

The discussion of the quiver gauge theory above is rather general,
and for practical purposes one needs to specify the theory more in detail.

First, we need to specify the spacetime dimension $D$ for the 
the quiver gauge theories are defined---for example, the gauge field $A_v$ introduced above
is $A_v^{\mu}(x)$, where $\mu$ runs from $0, \ldots, D-1$, and $x$ is a point of the $D$-dimensional geometry, such as $\mathbb{R}^{1,D-1}$.
In practice we will consider the case of $D=4,3,2,1$.

\begin{figure}[htbp]
\centering{\includegraphics[scale=0.4]{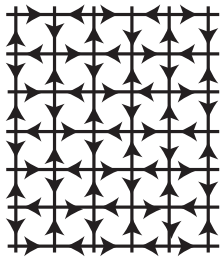}}
\caption{Statistical lattice where a statistical-mechanical model is defined.}
\label{Fig2_2}
\end{figure}

Second, we did not specify the representation of the matter field $\phi_e$ under the symmetry $G_v\times G_{v'}$
(where two endpoints of $e$ are denoted by $v, v' \in e$). In this lecture we choose 
the gauge group $G_v$ to be an $SU(N_v)$ gauge group (with rank specified by an integer $N_v$),
and the matter field to be in the so-called bifundamental representation,
namely in the representation $(\square, \overline{\square})$ under $G_v\times G_{v'}$. We then need to distinguish the two endpoints,\footnote{This is because the fundamental $\square$ and the anti-fundamental representation $\overline{\square}$ are different representation for $\SU(N)$ groups.
The exception happens for $N=2$, where the fundamental and anti-fundamental representations are isomorphic.}
and hence combinatorially we now need to consider oriented edges, as in Figure \ref{Fig2_2}.

That we need to specify the orientation of the arrow means that the our quiver gauge theory is now chiral.
This is indeed the case for almost all the theories we discuss in this paper (except in section \ref{sec.SU2}). In terms of statistical-mechanical models,
this chirality means that the Boltzmann weight for an edge $e$
is asymmetric with respect to the exchange of the endpoints of the edge:
\begin{align}
\mathcal{E}^e(s_{v}, s_{v'}) \ne \mathcal{E}^e(s_{v'}, s_{v}) \;.
\end{align}

In Figure \ref{Fig2_2} we have chosen the orientation of the edges such that 
for each face of the lattice (e.g.\ a square in Figure \ref{Fig2_2}) all the edges around the face are all in the same direction,
either all clockwise or all counterclockwise. This also means that we have a corresponding gauge-invariant product of the
bifundamentals around the face, which we can use as a term in the superpotential:
\be
W=\textrm{Tr}\left[\overset{\curvearrowright}{\prod}_{e\in \textrm{ clockwise face}}  \Phi_e  \right]-\textrm{Tr}\left[\overset{\curvearrowleft}{\prod}_{e\in \textrm{ counterclockwise face}} \Phi_e  \right] \;.
\label{W_rule}
\ee
This will be our general rule in the rest of this lecture.\footnote{This type of rules has been most systematically studied in the context of ``brane tilings'' \cite{Hanany:2005ve,Franco:2005rj}, see e.g.\ \cite{Kennaway:2007tq,Yamazaki:2008bt} for review.}\footnote{We have set the coefficients of the superpotential terms to be $1$. See \cite{Imamura:2007dc} for detailed discussion of the coefficients.}

Finally, we have not the boundary condition for the statistical-mechanical model: 
this can be say either periodic boundary condition or fixed boundary condition.
In quiver gauge theory language, the difference is whether to consider the quiver diagram on the torus \cite{Hanany:2005ve,Franco:2005rj}
or on a disc  \cite{Franco:2012mm,Xie:2012mr}. For the latter case we have external quiver vertices, whose associated symmetry
we regard as a non-dynamical flavor symmetry.

\subsection{Seiberg(-like) Duality}

\begin{figure}[htbp]
\centering{\includegraphics[scale=0.32]{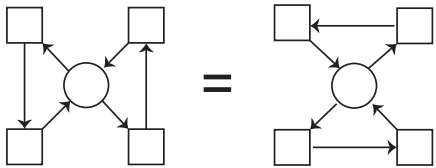}}
\caption{Graphical representation of the Seiberg(-like) duality for quiver gauge theories.
The same picture can also represent a quiver mutation in cluster algebras, or 
the star--star relation in integrable models.}
\label{Fig3_1}
\end{figure}

The final ingredient is the Seiberg(-like) duality.
This is represented graphically in Figure \ref{Fig3_1}.

The two quivers in Figure \ref{Fig3_1} can be regarded as the defining data for the quiver gauge theory.
As for the vertices, the circles represent the gauge symmetry, where as the squares represent the flavor symmetry.
Here and in the following we choose symmetry group to be $G_v=\SU(N)$ at all the vertices $v\in V$, either gauge or flavor,
and fix this integer $N$ throughout.\footnote{Seiberg duality in itself is known for other gauge groups, however its relevance for the Gauge/YBE correspondence is unknown as of this writing.}

The theory is known as the $\SU(N)$ SQCD with $2N$ flavors (since two arrows comes into the middle circle vertex,
and two arrows out).
The non-Abelian part of the flavor symmetry is $\SU(2N)_L\times \SU(2N)_R$,\footnote{There is also a $U(1)$ R-symmetry, whose role will be discussed in section \ref{sec.spectral}.}
which in this case is broken to its subgroup $\SU(N)^4$.
Such a breaking is caused by the presence of ``mesons'', which are represented as two 
arrow connecting a square to another, and hence transforms as bifundamental representation
with respect to the $\SU(N)\times \SU(N) \subset \SU(N)^4$ flavor symmetry.

For the case of 4d $\CN=1$ theories,
the duality between the two theories, which holds in the long distance limit,
is then the famous Seiberg duality  \cite{Seiberg:1994pq},
where the case of $N_f=2N$ flavors is right in the middle of the conformal window.
Note that following the previous rule \eqref{W_rule} we here include a cubic superpotential term for each triangle in Figure \ref{Fig3_1}, and such a superpotential term (meson--quark--quark superpotential term) is needed for 
the Seiberg duality.

There are similar versions in 3d $\CN=2$ \cite{Aharony:1997gp} and 2d $\CN=(2,2)$ \cite{Gadde:2013dda,Benini:2014mia},\footnote{In two dimensions
the gauge group at the vertex should be $\U(N)$, not $\SU(N)$.} and we will collectively refer to all of them as Seiberg-like dualities, or 
simply Seiberg dualities.

For mathematically-inclined readers Figure \ref{Fig3_1} can be thought of as a special example of the 
quiver mutation \cite{FominZelevinsky1,FominZelevinsky4}.
We will not really take advantage of this interpretation, partly because Seiberg duality in four dimensions corresponds only to a special subset of the 
more general quiver mutation (in two dimensions there is a closer relation between Seiberg duality and quiver mutation \cite{Benini:2014mia}, a fact which we will comment again later in this paper). 

It is fair to say that these two interpretations mentioned above of Figure \ref{Fig3_1}, in terms of Seiberg duality and quiver mutation,
are well-known in the literature.

What is less known is that the same graph has 
yet another interpretation in statistical mechanics---the star--star relation 
(SSR) \cite{Baxter:1986,Bazhanov:1992jqa}.\footnote{Interestingly, Baxter and Bazhanov recognized this relation years before the 
discovery of Seiberg duality. The connection to Seiberg duality, however, was noticed only recently.}

The motivation for SSR in the context of integrable models is that
SSR implies YBE, and hence once we solve the SSR we automatically
obtain an integrable model. This fact has far-reaching implications, and is one of the cornerstones of the Gauge/YBE correspondence.

We can now put together all the ingredients discussed in this section.

We can start with Figure \ref{Fig3_1}, which we can regard as the quiver diagram
and associate quiver gauge theories.
Thanks to Seiberg duality, we know that the two theories are dual,
and hence their partition functions coincide in the IR. Since Figure \ref{Fig3_1} is simultaneously the SSR,
we have solved the SSR by computing the supersymmetric partition function of the 
quiver gauge theories: that such a partition function has statistical-mechanical interpretation was explained in 
section \ref{sec.2.1} and \ref{sec.2.2}.  This means that we have solved SSR, and hence have solved the YBE!

\section{YBE as Duality}

While the explanation given in the previous section is enough to understand minimally
how we obtain integrable models, it is instructive to 
look into the construction in more detail. Interestingly,
almost all the ingredients in integrable models have counterparts in 
quiver gauge theories, and we find a step-by-step parallel.

\subsection{Theory for R-matrix}

\begin{figure}[htbp]
\centering{\includegraphics[scale=0.28]{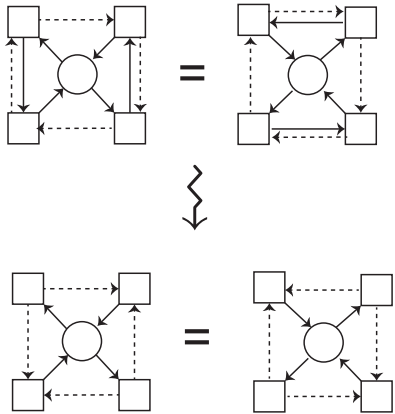}}
\caption{More symmetric representation of the SSR. Here a dotted line is a half-line, which when combined with a full-line in an opposite orientation
creates a half-line.}
\label{Fig3_2}
\end{figure}

For our later purposes it is useful to represent the SSR in Figure \ref{Fig3_1} more symmetrically
by adding half-arrows \cite{Yamazaki:2015voa}. This is shown in Figure \ref{Fig3_2},
where half-arrows are shown as dotted lines.

The rule is that when we have a full arrow in one direction and a half arrow
in the opposite direction, we can then cancel them into a half-arrow (as in Figure \ref{Fig3_2}). 
After this manipulation
we obtain an expression as in the bottom of Figure \ref{Fig3_2}.

One should note that there is no such thing as a half-matter (say a half-chiral multiplet) in quiver gauge theories.
Such a half-line, however, helps to explain the integrability structure in a more symmetric and unified manner.
Note that in our following discussions half-arrows always connect flavor symmetry gauge groups (squares),
and hence non-dynamical. At the level of the supersymmetric partition functions the only change from the full matter is that 
we take a square root of the corresponding classical/one-loop contribution. 
In any case, whenever we discuss dualities among quiver gauge theories we can always cancels the half arrows
so that the duality can be stated in terms of full arrows only (we will see this in Figure \ref{Fig8_4}.).

The claim now is that the quivers in Figure \ref{Fig3_2} should be regarded as
the quiver for the R-matrix (Figure \ref{Fig8_1}):

\begin{figure}[htbp]
\centering{\includegraphics[scale=0.33]{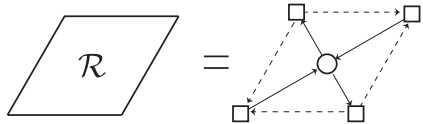}}
\caption{We associate he quiver in the Figure \ref{Fig3_2} to the R-matrix of the integrable model:
the R-matrix is now promoted to a ``theory'' $\mathcal{T}[R]$!}
\label{Fig8_1}
\end{figure}

In other words, the theory $\mathcal{T}[R]$ specified by the quiver\footnote{Since the quiver contains a half-arrow,
the ``theory'' $\mathcal{T}[R]$ is strictly speaking is not an authentic theory, as explained above. We need to add half-arrows appropriately to 
make it into an actual theory. It is nevertheless in practice convenient to refer to $\mathcal{T}[R]$ as a theory.} is the
``gauge-theory uplift'' of the R-matrix.
To better understand this statement we need to discuss the YBE for this R-matrix.

\subsection{Gluing as Gauging}

In order to discuss YBE we need to combine the three R-matrices. Let us see how this works
at the level of the quiver diagram, and hence of the quiver gauge theory.

\begin{figure}[htbp]
\centering{\includegraphics[scale=0.33]{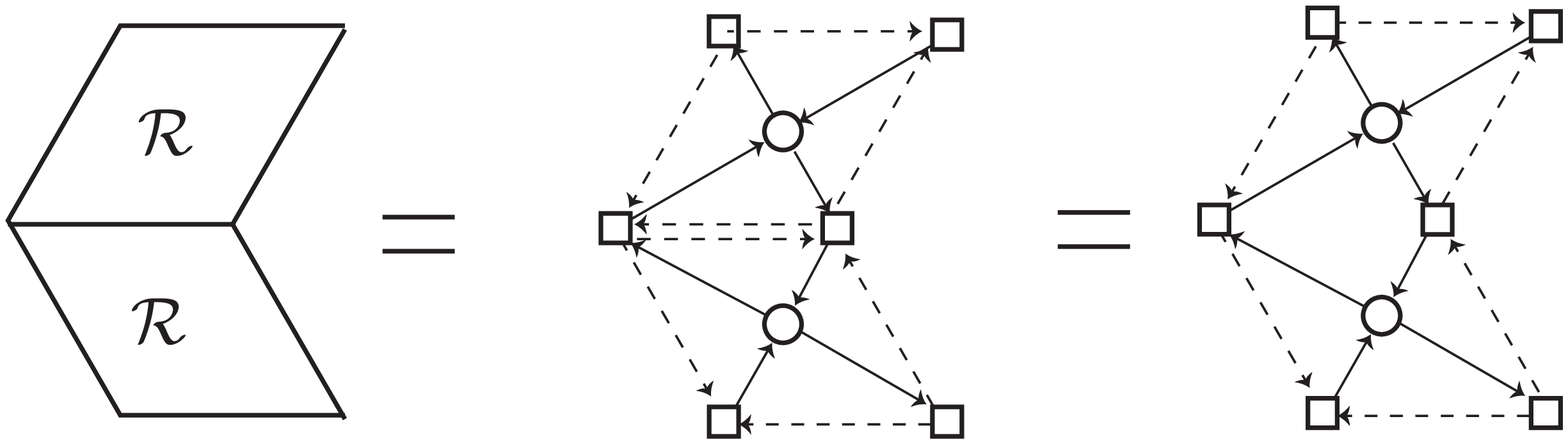}}
\caption{Gluing of two R-matrices. }
\label{Fig8_2}
\end{figure}

Let us first glue two R-matrices. In the left figure of Figure \ref{Fig8_2} we need to cancel the half-lines in the opposite directions,
to obtain the quiver as in the right of Figure \ref{Fig8_2}.

\begin{figure}[htbp]
\centering{\includegraphics[scale=0.33]{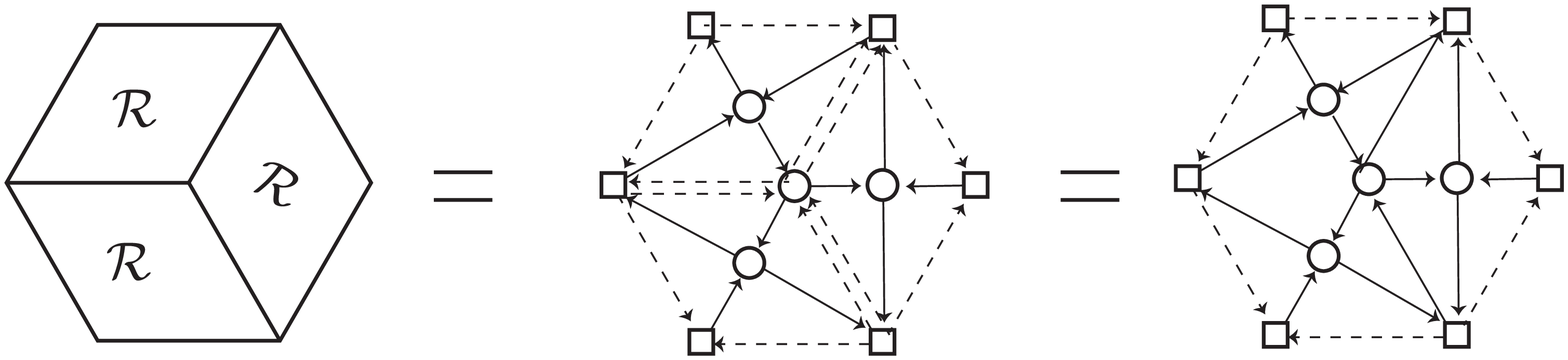}}
\caption{We combine the three R-matrices. Depending on the orientations the two half-arrows
either cancels or combines into a full arrow. The vertex in the center, now an internal vertex, is promoted to the gauge symmetry.}
\label{Fig8_3}
\end{figure}

We can next combine three R-matrices as in Figure \ref{Fig8_3}.
We again cancel the two half-arrows in opposite directions; in addition we combine
the two half-arrows into a one full arrow, when the two are in the same direction.

Moreover the vertex in the center of the figure at the contact point of three rhombi
is now turned into a circle, namely the corresponding $\SU(N)$ symmetry is gauged.
In general, the rule is that an internal vertex is gauged, whereas an external vertex is not gauged.\footnote{This rule in natural string theory, where the 
$\SU(N)$ symmetry arises from $N$ D-branes. The gauge coupling is finite if the areas wrapped by that D-branes is finite,
otherwise when the D-branes run off to infinity then the area is infinite and the gauge coupling constant is zero, thus making the symmetry non-dynamical.
See e.g.\ \cite{Heckman:2012jh} for a discussion for the type for theories discussed in this paper.}
In this sense we are gluing the three copies of the theory $\mathcal{T}[R]$ by appropriate gauging.

Note that the idea of  ``gluing the theories by gauging'' is has appeared in many physics problems in the past, and has been popular after the discovery of the so-called class $S$ theories \cite{Gaiotto:2009we}. Indeed, many of our discussions here is parallel to the discussion there. For example the role of the trinion theory (the so-called $T_N$ theory) there is played by our R-matrix theory $\mathcal{T}[R]$. The choice of the Riemann surface obtained by gluing trinion, is here played either by the choice of the toric diagram for quivers on the torus \cite{Hanany:2005ve,Franco:2005rj} and the choice of the decorated permutation \cite{Xie:2012mr,Franco:2012mm}\footnote{The decorated permutation labels the positroid cell of the positive Grassmannian,
which combinatorial structure was used intensively in the discussion of scattering amplitudes of 4d $\CN=4$ supersymmetric Yang-Mills theory \cite{ArkaniHamed:2012nw}. One can take this point more seriously and introduce a spectral parameter deformation to the scattering amplitude \cite{Ferro:2012xw,Ferro:2013dga,Bargheer:2014mxa,Ferro:2014gca}. From integrable model viewpoint, this gives a Grassmannian formula for a R-matrix of the Yangians for $\mathfrak{psu}(4|4)$ and $\mathfrak{osp}(6|4)$.}
for quivers on the disc.

\subsection{Yang-Baxter Duality}

\begin{figure}[htbp]
\centering{\includegraphics[scale=0.29]{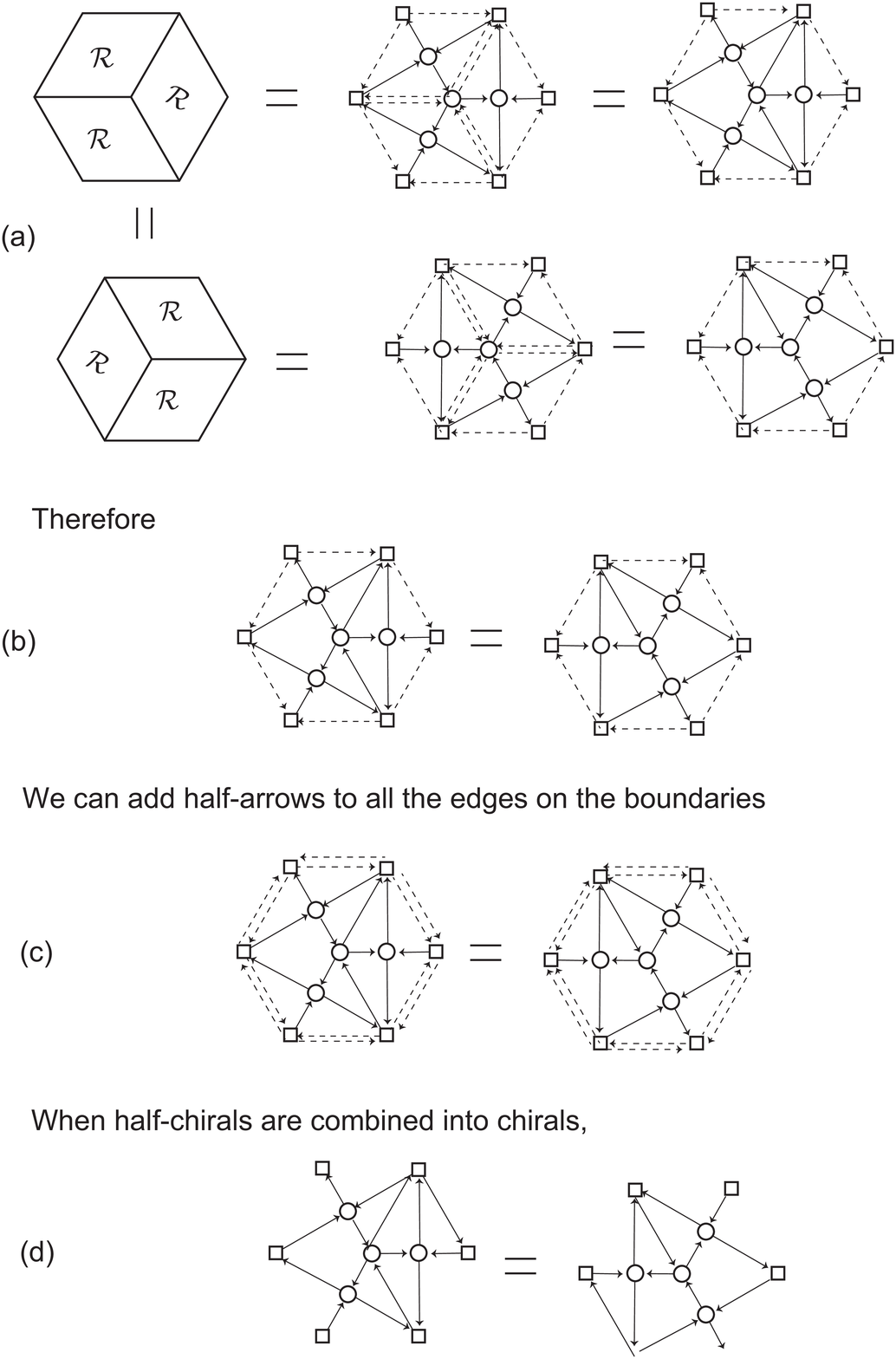}}
\caption{The quiver counterpart of the YBE. After combining the three R-matrices as in (a)(b). After adding the half-arrows for the external global symmetries as in (c) and canceling/combining the half-arrows on both sides of the identity, we obtain an equality in (d). The gauge theory interpretation of this is that the two quivers in (c) are dual, namely they flow to the same IR fixed point.}
\label{Fig8_4}
\end{figure}

We are now ready to discus the YBE. Since the combination of the three R-matrices shown in Figure \ref{Fig8_3}
is the left hand side of the YBE,\footnote{Here we are using the formulation of the integrable model as an IRF (interaction-around-a-face) model.
Our integrable model can also be formulated as a vertex model.} we can do the same for the right hand side and immediately write down the quiver counterpart of the YBE, to obtain Figure \ref{Fig8_4}(b). We can again add half-arrows as in  Figure \ref{Fig8_4}(c), and then cancel/combine the half-arrows, to 
obtain Figure \ref{Fig8_4}(d).

The final equality in Figure \ref{Fig8_4}(d) contains a precise gauge theory statement.
Namely, the quiver gauge theories, described by two quivers in Figure \ref{Fig8_4}(d), are dual, namely they flow to the same IR fixed point.
Since this duality is the precise counterpart of the YBE, let us call this duality {\it the Yang-Baxter duality}.\footnote{When I had the chance to talk with Prof.\ Rodney J.\ Baxter
in 2015, I introduced the terminology ``Yang-Baxter duality''. His immediate reaction was ``I'm not a duality!''. The terminology ``Yang-Baxter duality'' is not meant to be a personal duality between C.N.\ Yang and R.J.\ Baxter; as is hopefully clear the terminology was chosen since this is a duality underlying YBE.}

\begin{figure}[htbp]
\centering{\includegraphics[scale=0.43]{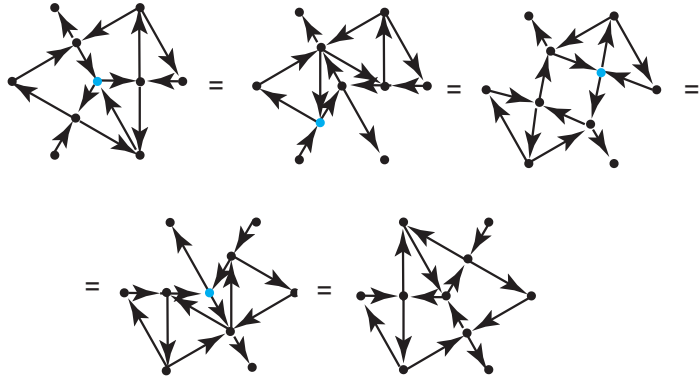}}
\caption{The Yang-Baxter duality follows by repeated use of the Seiberg duality.
In this figure the blue vertices denote the gauge groups where we perform the Seiberg duality (quiver mutation)
in the next step.}
\label{Fig8_5}
\end{figure}

Of course, the question is then if this duality holds. Fortunately for us, we can show that the Yang-Baxter duality follows
by repeated use of the Seiberg duality, as shown in Figure \ref{Fig8_5}.

Of course, this is essentially the explanation that the SSR implies the YBE, a fact
which was known already in the old literature of integrable models \cite{Baxter:1986,Bazhanov:1992jqa}.
What is new here is that we are now discussing the quiver gauge theories attached to the same figures,
and as we will see this understanding is remarkably powerful in constructing new integrable models.

\section{Spectral Parameter as R-charge}\label{sec.spectral}

\subsection{Zig-Zag Paths and R-charge}

There is still one missing ingredient in our discussion so far: the spectral parameter 
in integrable models. As commented before, this parameter is essential in obtaining an infinitely-many conserved charges.

In the Gauge/YBE correspondence, the gauge-theory counterpart of the spectral parameter in integrable models
is the R-charge. Simply put, in both cases the parameters are associated with the so-called ``zig-zag paths''.\footnote{
The relevance of zig-zag paths for quiver gauge theories was pointed out in \cite{Hanany:2005ss}. Zig-zag paths are also discussed in the mathematical literature on bipartite graphs, see e.g. \cite{ThurstonDomino,Goncharov:2011hp}.}

\begin{figure}[htbp]
\centering{\includegraphics[scale=0.22]{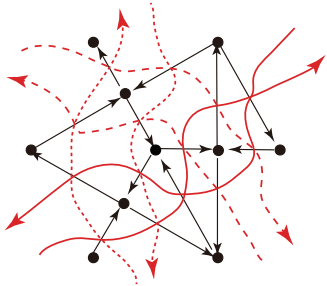}}
\caption{Zig-zag paths, represented by red lines. We have shown the zig-zag paths as dotted or undotted lines,
to make it easier to distinguish between different lines. We can recover the original quiver diagram from the zig-zag paths only.
For this reason the zig-zag paths contain the same data as the original quiver diagram.}
\label{Fig5}
\end{figure}

Since a picture is worth a thousand words, the best way to explain the zig-zag path is simply to show Figure \ref{Fig5},
and I hope that the rule is self-explanatory. In Figure \ref{Fig5} the zig-zag paths (colored red) is a path coming in from infinity
and eventually goes off to infinity. They intersect precisely once for each edge of the original quiver diagram (colored black).

The zig-zag paths give a nice parametrizations of the R-charges.
Let us label the zig-zag paths by $i=1,2,\ldots$, and let us associate
an angle $\theta_i$ for each path. Since $\theta_i$ is an angle, we have the identification $\theta_i\sim \theta_i+2\pi$.
Suppose that one wants to know the R-charge of the matter multiplet. 
In quiver gauge theory such a matter is located at an edge of the quiver diagram,
where two zig-zag paths, say $i$ and $j$, cross.
Then the R-charge of the matter multiplet is 
given by $\theta_{ij} / \pi$, where  $\theta_{ij}$ is the relative angle between $\theta_i$ and $\theta_j$.
Readers are referred to \cite{Hanany:2005ss,Yamazaki:2012cp,Xie:2012mr}
for more details of this parametrization. Note that the condition that the sum of R-charges is $2$ is analogous to the statement that the 
sum of the R-charges is $2\pi$, and the point is that one can promote this analogy into a precise correspondence.\footnote{The idea that the R-charge can be identified with an angle 
also appeared in the context of the 3d--3d correspondence \cite{Terashima:2011qi,Terashima:2011xe,Dimofte:2011jd,Dimofte:2011ju,Dimofte:2011py,Cecotti:2011iy}. The relation between Gauge/YBE and 3d--3d deserves further exploration, see \cite{Yamazaki:2012cp,Terashima:2012cx,Yagi:2015lha} for 
preliminary discussion.}\footnote{The R-charge in itself is a real parameter. However, we can naturally complexify it into a complex parameter,
see \cite{Festuccia:2011ws} for explanation from supergravity.}

\begin{figure}[htbp]
\centering{\includegraphics[scale=0.17]{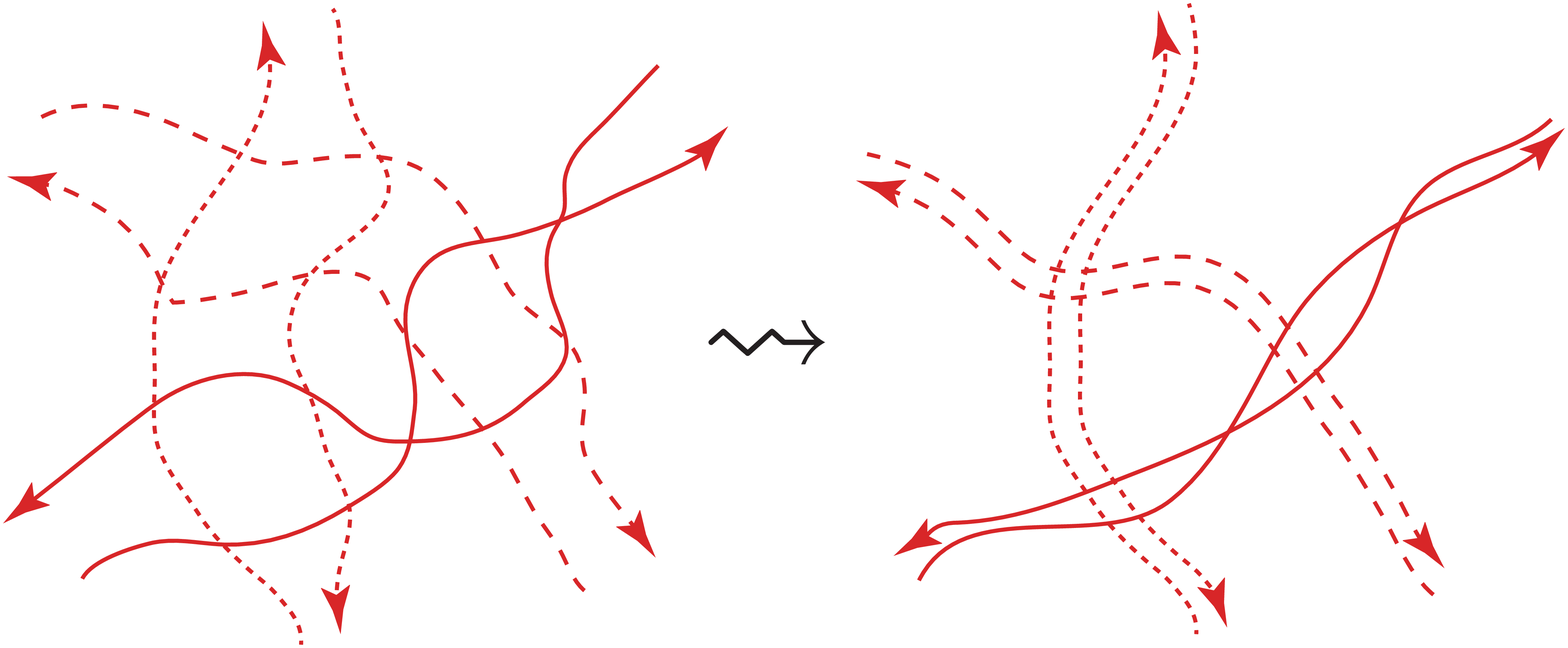}}
\caption{We can disregard the quiver diagram from Figure \ref{Fig5}, to obtain a set of zig-zag paths as in this figure.
By appropriately deforming the paths (while keeping the topology of the graph) we arrive at the picture on the right,
which is very similar to the picture for the YBE (recall Figure \ref{Fig1}). }
\label{Fig6}
\end{figure}

\begin{figure}[htbp]
\centering{\includegraphics[scale=0.23]{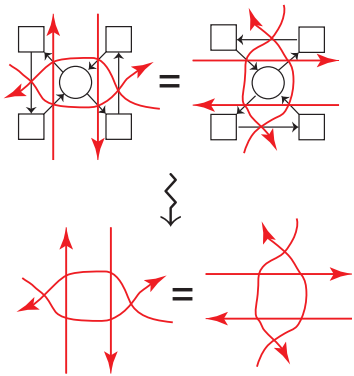}}
\caption{Graphical representation of SSR, now in terms of the zig-zag paths.
SSR is here represented by a move involving four zig-zag paths.}
\label{Fig10}
\end{figure}

\begin{figure}[htbp]
\centering{\includegraphics[scale=0.17]{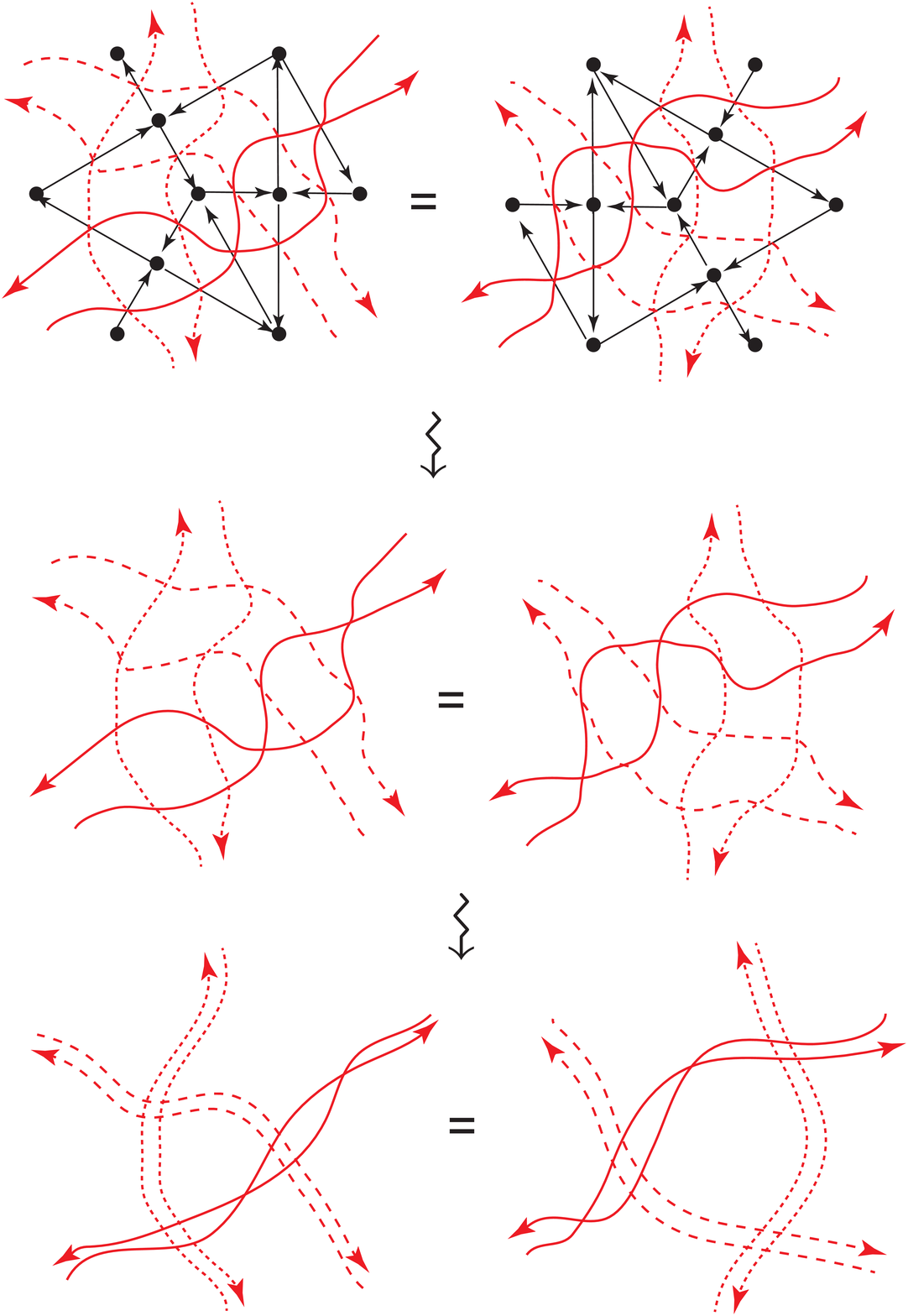}}
\caption{YBE as implied by SSR, now reformulated in terms of zig-zag paths. The resulting picture has doubled rapidity lines running parallel in the opposite  directions, and has `twists' in several places.}
\label{Fig7}
\end{figure}

Apart from parametrizing R-charges, the zig-zag paths contain exactly the same data as the original quiver diagram.
In fact, by looking at Figure \ref{Fig5} one can convince oneself that we can recover the original quiver diagram from the zig-zag paths only.
For this reason, we are free to disregard the quiver diagram and keep only the zig-zag paths.\footnote{This is a variant of the Baxter's Z-invariant lattice \cite{Baxter:1978xr}.}
If we do this for the left figure of Figure \ref{Fig6} and deform the paths (while keeping the topology of the graph),
we arrive at the right figure of Figure \ref{Fig6}, which is very similar to the picture for the YBE (recall Figure \ref{Fig1}). 
We can also represent the SSR and the YBE in terms of the zig-zag paths, as in Figures \ref{Fig10} and \ref{Fig7}.
The previous statement that SSR implies YBE is now an exercise in graphical calculus: we need to show the equality of the zig-zag paths in Figure \ref{Fig7}
by using the moves in Figure \ref{Fig10}. I would like to invite the readers to check this statement directly to one's satisfaction.

There are still some differences from Figure \ref{Fig1}. First, the role of a single rapidity line is now
played by a pair of zig-zag paths which run parallel with each other in the opposite direction.\footnote{By using the parametrization of the zig-zag paths 
we actually find a $2$-parameter extension of the R-matrix (in addition to the spectral parameter),
and the Gauge/YBE correspondence generates a solution to a certain generalization of the YBE with these extra parameters, see \cite{Yamazaki:2015voa}.} Second, in the bottom figure of Figure \ref{Fig7} such a pair of zig-zag paths are 
twisted in four locations. One can therefore say what we obtain here is a ``doubled twisted'' YBE.

The existence of the `twist' of course does not mean that 
we have failed to solve the YBE. Recall that whole construction arises from
solving the YBE as in Figure \ref{Fig8_4}, and we have already seen in Figure \ref{Fig8_5}
that the Seiberg-like duality (solution to SSR) solves the YBE on the nose.
The `twist' in Figure \ref{Fig7} arises only when we try to eliminate the half-arrows and represent everything in terms of zig-zag paths.

The origin of this `twist'  can be traced back to the existence of the mesons  in the Seiberg duality in Figure \ref{Fig3_2}.
It is interesting to see that the integrable model experts, at least in their own context, knew that mesons are needed for a 
proper discussion of SSR (Seiberg-like duality), years before the discovery of Seiberg duality in 1994.

\subsection{Zig-Zag Paths from String Theory}

We have so far introduced the zig-zag paths as purely combinatorial objects. This is often the case in the literature,
either in integrable models or in quiver gauge theories. Fortunately we can do better in string theory
and identify the zig-zag paths as the ``shape of the branes''.

Let us discuss this point briefly for the case of 4d $\mathcal{N}=1$ quiver gauge theories, following the author's Master's thesis \cite{Yamazaki:2008bt}.\footnote{See also \cite{Imamura:2006ie,Imamura:2007dc} for related earlier work. The type IIB brane configuration here is  mirror to type IIA description of \cite{Feng:2005gw}. Our discussion here is when the quiver diagram is written on the torus; see e.g.\ \cite{Heckman:2012jh} for quiver diagrams on the disc.}

Let us consider type IIB string theory on 
$\mathbb{R}^{3,1}_{0123}\times (\mathbb{C}^{\times})^2_{4567}\times \mathbb{C}_{89}$.
$N$ D5-brane spread in $012357$-directions,
with $57$ directions compactified (namely gives $T^2$), realizing $\SU(N)$ gauge groups. 
We in addition have one NS5-brane, 
filling the $0123$ directions as well as 
a holomorphic curve $\Sigma(x,y)=0$ (with $x=e^{x_4+i x_5}, y=e^{x_6+i x_7}$)
inside $(\mathbb{C}^{\times})^2$ ($4567$-directions), see Table \ref{tbl:D5NS5}.

\begin{table}[htpb]
\caption{The type IIB configuration realizing 4d $\mathcal{N}=1$ quiver gauge theories.}
\begin{center}
\begin{tabular}{c|cccc|cccc|cc}
\hline
\hline
&0&1&2&3&4&5&6&7&8&9 \\
\hline
D5&$\circ$ &$\circ$ &$\circ$ &$\circ$ & & $\circ$ & &$\circ$& & \\
NS5&$\circ$ &$\circ$ &$\circ$ &$\circ$ &\multicolumn{4}{c}{$\Sigma$ (2-dim surface)} & \multicolumn{1}{|c}{} \\
\hline
\end{tabular}
\label{tbl:D5NS5}
\end{center}

\end{table}

The NS5-brane and D5-brane intersect along paths (1-cycles) on the $57$-torus,
and these paths are identified with the zig-zag paths introduced before. The zig-zag paths divide the torus into various regions, realizing the quiver structure of the gauge group (see \cite{Yamazaki:2008bt} for details).\footnote{A similar setup has been discussed more recently in the mathematical work of \cite{Shende:2015mzx}.}
The Seiberg duality (as in Figure \ref{Fig10}) can be understood as a reordering of the five-branes reminiscent of the Hanany--Witten transition \cite{Hanany:1996ie}. Note that in this brane setup 
the R-symmetry (which we associate with the spectral parameters) can be identified with the rotation symmetry in the transverse $89$-plane.

Other than providing a natural explanation of the zig-zag paths, the five-brane configuration
could be a useful starting point for exploring other approaches to integrable models,
for example with the ``4d Chern-Simons'' approach mentioned in introduction. This requires further study, see e.g. \cite{Yagi:2015lha,Ashwinkumar:2018tmm,Costello:2018txb,Yamazaki:2019prm}.
 
\section{\texorpdfstring{$\SU(2)$ and Star-Triangle Relation}{\SU(2) and Star-Triangle Relation}}\label{sec.SU2}

Let us next comment on the special case where
the gauge groups at the quiver vertex is $\SU(2)$ (as oppose to $\SU(N)$ with $N\ge 2$).
This is special since in this case there is no distinction between the fundamental representation and the anti-fundamental representation.
Correspondingly the theory is non-chiral, and there is no need to specify the orientation of the edges of the quiver diagram,
and of the zig-zag paths.

\begin{figure}[htbp]
\centering{\includegraphics[scale=0.25]{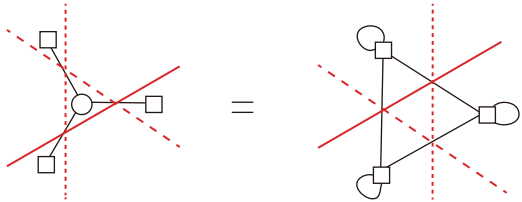}}
\caption{For $\SU(2)$ gauge groups we can consider the duality for the star-triangle relation.
This in turn implies the SSR, and hence is more fundamental as a duality. In terms of zig-zag paths
this represents the reshuffling of the positions of the three zig-zag paths. Note that the zig-zag paths are here not oriented, and the theory is non-chiral.}
\label{Fig9}
\end{figure}

Of course, everything we said so far applies to the special case of $\SU(2)$. The notable difference for the $\SU(2)$ case
is that the Seiberg duality as represented in Figure \ref{Fig3_2} follows in turn from another 
Seiberg duality, as shown in Figure \ref{Fig9}.

In the left figure of Figure \ref{Fig9} the gauge vertex in the middle 
has three arrows coming in, and hence we have $N_f=3$ (namely $6$ doublets),
with $\SU(6)$ symmetry broken into $\SU(2)^3$.
This theory is known (again by the Seiberg duality \cite{Seiberg:1994pq}) to a non-gauge theory with $\binom{6}{2}=15$ mesons.
In the right figure of Figure \ref{Fig9}, $12$ of these mesons are shown as edges connecting two flavor $\SU(2)$ symmetries, where the remaining $3$ mesons are in the adjoint representation with respect to the 
one of the $\SU(2)$ flavor symmetries.

Figure \ref{Fig9} is known in integrable model literature as the star-triangle relation (STR), one of the expressions for the YBE \cite{Baxter:1982zz}.
We therefore conclude that the duality in Figure \ref{Fig9} generates a solution to the STR \cite{Spiridonov:2010em}.\footnote{The three adjoints are not charged under the gauge symmetry and hence appear as the normalization factor outside the integral. Such a normalization factor has been considered in the 
general discussion of the STR.}

In terms of zig-zag paths 
Figure \ref{Fig9} means that we can move the relative positions of any three zig-zag paths. This implies the SSR and then in turn the YBE (representation of the Yang-Baxter duality) for our R-matrix; in terms of zig-zag paths both are represented as certain moves of the paths (see Figures \ref{Fig10} and \ref{Fig7}), and these easily  follow since we can change the position of any three paths, and hence of any paths.

Summarizing, we find
\begin{equation}
\begin{tabular}{c}
\textrm{STR: ($\SU(2)$ Seiberg duality with $N_f=3$ ($6$ doublets))}\\
$\downarrow$ \\
\textrm{SSR: ($\SU(2)$ Seiberg duality with $N_f=4$ ($8$ doublets))}\\
$\downarrow$\\ 
\textrm{YBE: (Yang-Baxter duality)} \;.
\end{tabular}
\end{equation}

\section{New Integrable Models}

Having explained the general theory,
let us now turn to the question: which concrete integrable model
arises from the general construction of the previous section?

\subsection{General Lessons}

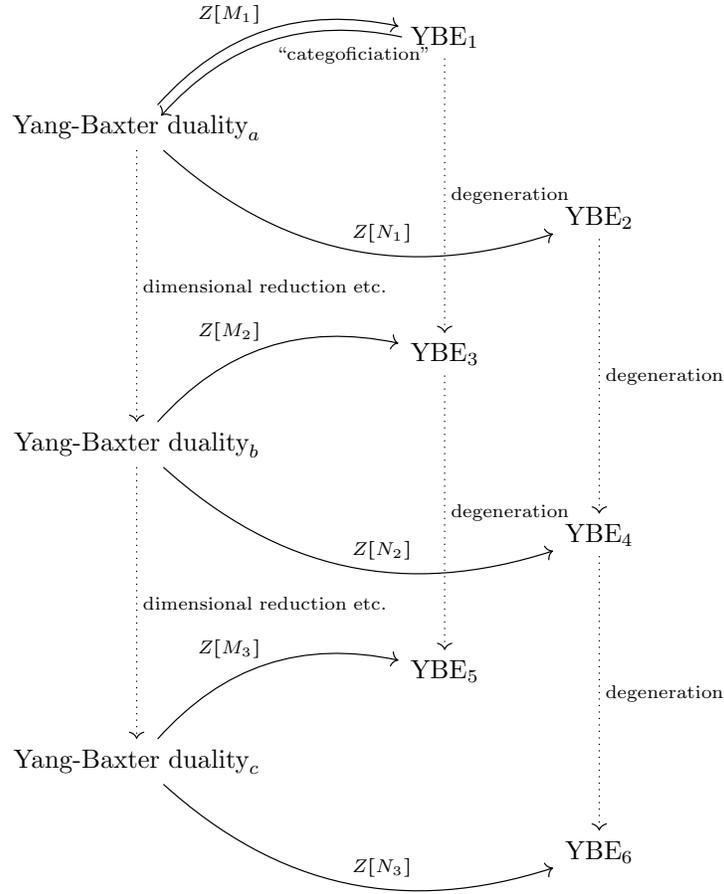
\begin{figure}[htbp]
\begin{equation*}
\begin{tikzcd}
 & & \textrm{YBE}_1 \arrow[dll,bend right,shift left=1ex,"\textrm{``categoficiation''}"] 
 \arrow[dddd, dotted,"\textrm{degeneration}"]&  \\
 \textrm{Yang-Baxter duality}_a \arrow[dddd,dotted, "\textrm{dimensional reduction etc.}"] \arrow[urr,bend left,"Z {[ M_1]}"] \arrow[drrr,bend right,"Z {[ N_1]}"] & &  &  \\
  && &  \textrm{YBE}_2  \arrow[dddd, dotted,"\textrm{degeneration}"]\\  
 & & &   \\
  &  & \textrm{YBE}_3\arrow[dddd, dotted,"\textrm{degeneration}"] &  \\
  \textrm{Yang-Baxter duality}_b \arrow[dddd,dotted, "\textrm{dimensional reduction etc.}"]\arrow[urr,bend left, "Z {[ M_2]}"] \arrow[drrr,bend right,"Z {[ N_2]}"] & &  &  \\
 & & &  \textrm{YBE}_4 \arrow[dddd, dotted,"\textrm{degeneration}"]\\  
  & & &   \\
  &  & \textrm{YBE}_5 &  \\
  \textrm{Yang-Baxter duality}_c \arrow[urr,bend left, "Z {[ M_3]}"] \arrow[drrr,bend right,"Z {[ N_3]}"] & &  &  \\
 & & &  \textrm{YBE}_6 \\  
\end{tikzcd}
\end{equation*}
\caption{The Yang-Baxter duality ``uplifts'' integrable models into gauge theories and 
unifies many integrable models which are unrelated otherwise. Namely, starting with a single duality we can 
compute various different supersymmetric partition functions and obtain
many different dualities. In addition we can change the gauge theory side by for example
dimensionally reducing the theory. Such a gauge theory operation, as long as it preserves the duality,
has counterparts in the integrable model side.}
\label{Fig.overall}
\end{figure}

Before answering this question, 
let us first summarize our overall logic as in Figure \ref{Fig.overall}.

In a typical discussion of integrable models we choose a specific integrable model (a solution of YBE),
for example the one denoted as YBE$_1$ in Figure \ref{Fig.overall}. Rather, here we 
are interested in the underlying Yang-Baxter duality (denoted Yang-Baxter duality$_a$ in  Figure \ref{Fig.overall}):
this can be thought of as a ``gauge-theory uplift''\footnote{Or a ``categorification'' in a rather loose sense.} of YBE$_1$, and conversely YBE$_1$ is obtained by computing a 
supersymmetric partition function $Z[M_1]$ on a manifold $M_1$.

Once we have the Yang-Baxter duality we can systematically construct a variety of integrable models,
by computing a different supersymmetric partition function (for example $Z[N_1]$ in Figure \ref{Fig.overall}).
In this sense the Yang-Baxter duality unifies many different integrable models, which as integrable models
look completely unrelated.

Once we establish the correspondence we can further generalize the setup by 
applying various operations on the field theory side. For example, we can dimensionally reduce the theory
along some directions of the compactification manifold, or we can integrate out some matters by giving mass.
As long as these operations preserve the duality\footnote{Since the Yang-Baxter duality is an IR duality, it is a non-trivial question
whether flowing to the IR and dimensional reduction commute. See \cite{Aharony:2013dha} for a related discussion. For our practical purposes we do not necessarily have to go through all the intricacies of the reduction, and could start with the known lower-dimensional version of the Seiberg and Yang-Baxter dualities. The degeneration at the level of the integrable models can then be verified independently.}, one should land on another 
version of the Yang-Baxter duality, which we called Yang-Baxter duality$_b$ in Figure \ref{Fig.overall}.

The relation between two different versions of the Yang-Baxter duality (Yang-Baxter duality$_a$ and Yang-Baxter duality$_b$ in Figure \ref{Fig.overall}) on the field theory side can now be translated into the 
integrable model side. For example, suppose that we dimensionally reduced the field theory along a circle, and suppose that the manifold $M_1$ contains a circle as a fiber, with the based manifold $M_2$. We then expect that the partition function on $M_1$ should reduce to the partition function on $M_2$,
and this process looks like a degeneration of the associated integrable models.

We can repeat the procedure above, and obtain a whole zoo of integrable models.

\subsection{New Integrable Models}

Let us now make the setup more concrete. We first need to specify which 
spacetime dimension we are in (in all the dimensions we consider a supersymmetric theory with four supercharges).
We then need to choose a supersymmetric partition function, which we represent by the choice of the manifold $M$.\footnote{In general even if we specify the geometry $M$ there are still choices to made as to which supergravity backgrounds we consider, and in general different such choices might lead to different answers of supersymmetric partition functions. For our purposes at hand it is in practice sufficient to specify the geometry only,
and hence we use the manifold $M$ itself to represent the choice of the supersymmetric partition function.} 

Once we fix the choice of the supersymmetric partition function, the resulting partition function is 
in general a function of the certain data of the manifold $M$. For example, 
for $S^1\times S^3$ the partition function depends on two continuous parameter $\p$ and $\q$,
which parametrizes the complex structure on $S^1\times S^3$ \cite{Kodaira,Closset:2013vra}.
The partition function is (up to the overall supersymmetric Casimir energy contribution)
the same as the superconformal index of \cite{Kinney:2005ej}, where the parameters $\p, \q$ are the fugacities in the definitions of the index. 

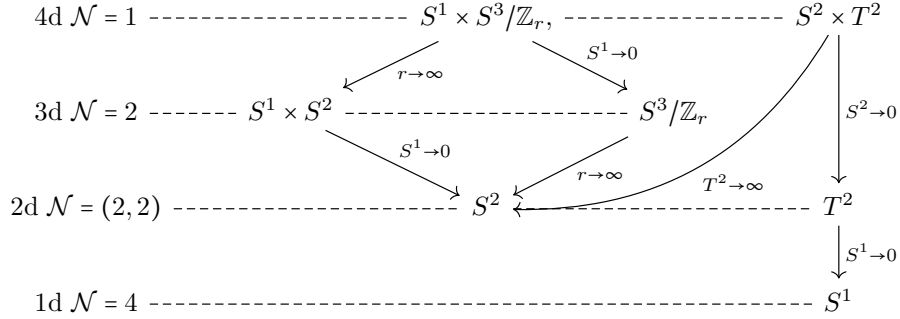
\begin{figure}[htbp]
\begin{equation*}
\begin{tikzcd}
 \textrm{4d $\CN=1$}  \arrow[rr,dashed,dash]&& S^1\times S^3/\mathbb{Z}_r, \arrow[dl,"r\to\infty"]\arrow[rr,dashed,dash] \arrow[dr,"S^1\to 0"]& & S^2\times T^2\arrow[ddll,bend left,"T^2\to\infty"] \arrow[dd,"S^2\to 0"]  \\
 \textrm{3d $\CN=2$}  \arrow[r,dashed,dash]&S^1\times S^2\arrow[dr,"S^1\to 0"]   \arrow[rr,dashed,dash] & & S^3/\mathbb{Z}_r \arrow[dl,"r\to\infty"]  &\\
 \textrm{2d $\CN=(2,2)$} \arrow[rr,dashed,dash]&   & S^2 \arrow[rr,dashed,dash]&   &T^2\arrow[d, "S^1\to 0"]\\
  \textrm{1d $\CN=4$} \arrow[rrrr,dashed,dash]&   &  && S^1  \\
\end{tikzcd}
\end{equation*}
\caption{A zoo of supersymmetric partition functions (and of associated integrable models). The arrows represents a dimensional reduction into one 
dimension lower. The case of $S^1\times S^3/\mathbb{Z}_r$ in four dimensions, whose associated integrable model is the super-master solution of YBE, is particularly general, and contains many other known solutions in the literature.}
\label{Fig.zoo}
\end{figure}

The actual integrable model has been discussed in many papers in the literature. While it is difficult to precisely summarize the rich literature,
it might be useful for readers to loosely classify the literature into: 4d $\mathcal{N}=1$ on $S^1\times S^3$ \cite{Bazhanov:2010kz,Bazhanov:2011mz,Spiridonov:2010em,Yamazaki:2012cp,Terashima:2012cx,Xie:2012mr,Maruyoshi:2016caf,Bazhanov:2016ajm,Yagi:2017hmj}, on $S^1\times S^3/\mathbb{Z}_r$ \cite{Yamazaki:2013nra,Spiridonov:2016uae,Kels:2015bda,Gahramanov:2016ilb,Kels:2017toi},
3d $\mathcal{N}=2$ on $S^1\times S^2$ \cite{Yagi:2015lha,Gahramanov:2015cva,Gahramanov:2016sen,Gahramanov:2017idz}, on $S^3$ \cite{Bazhanov:2007mh,Yamazaki:2012cp,Terashima:2012cx} and $S^3/\mathbb{Z}_r$ \cite{Gahramanov:2016ilb},
2d $\mathcal{N}=(2,2)$ on $S^2$ \cite{Yamazaki:2016wnu}, on $T^2$ \cite{Yagi:2015lha,Yamazaki:2015voa,Jafarzade:2017fsc}, and 1d $\mathcal{N}=4$ on $S^1$ \cite{Yamazaki:2016wnu}.

Rather than listing all the solutions, let us here comment on some notable features of the 
resulting integrable models.

The Boltzmann weights are written in terms of some special functions.
For example,
\begin{itemize}

\item For 4d $S^1\times S^3/\mathbb{Z}_r$ we find the elliptic gamma function (as noticed first in \cite{Dolan:2008qi})\footnote{For $r>1$ it is natural to define a certain combination of the elliptic gamma function, known as the lens elliptic gamma function \cite{Yamazaki:2013nra,Kels:2015bda,Kels:2017toi,Spiridonov:2016uae}. It would be interesting to further study the properties of this special function.}
\be
\Gamma(x;\p,\q)=\prod_{j,k\ge 0} \frac{1-x^{-1} \p^{j+1} \q^{k+1}} {1-x \p^j \q^j} \;, \quad |\p|, |\q|<1 \;.
\label{ell_gamma}
\ee

\item For 3d $\mathcal{N}=2$ on $S^1\times S^2$, on $S^3/\mathbb{Z}_r$
we have the $\q$-Pochhammer symbol\footnote{For $S^3/\mathbb{Z}_r$ we also have the quantum dilogarithm function,
which is written as a certain ratio of the Pochhammer symbols.}
\be
(x;\q)=\prod_{j=0}^{\infty} (1-x\q^j) \;, \quad |\q|<1 \;.
\ee

\item For 2d $\CN=(2,2)$ theory on $S^2$ we have the gamma function
\be
\Gamma(x)=\frac{e^{-\gamma x}}{x}\prod_{n=1}^{\infty} 
\frac{e^{\frac{x}{n}}}{\left( 1+\frac{x}{n} \right)} \;,
\ee
where $\gamma$ is the Euler--Mascheroni constant.

\end{itemize}

That the theories in 4d, 3d and 2d are related by dimensional reduction is reflected in the following hierarchy of degenerations
in the special functions:\cite{Gadde:2011ia,Dolan:2011rp,Imamura:2011uw,Benini:2012ui,Yamazaki:2013fva}
\be
\Gamma(x;\p, \q) \longrightarrow (x;\q) \longrightarrow \Gamma(x) \;.
\ee
This can be regarded as the elliptic--trigonometric--rational hierarchy of integrable models,
somewhat reminiscent of the Belavin--Drinfeld classification \cite{BelavinDrinfeld} of integrable models.
One should note, however, almost all the integrable models constructed from the Gauge/YBE correspondence does not 
fit into the classification scheme of \cite{BelavinDrinfeld}. In \cite{BelavinDrinfeld} the R-matrix is quasi-classical, namely
has a perturbative expansion starting with identity. Our models, however, do not seem to have such a property.\footnote{One can of course divide our R-matrix by some normalization factor, to find a perturbative expansion starting with identity. However, the normalization in itself will then be written as an complicated integral, which looks highly unnatural. Moreover, \cite{BelavinDrinfeld} assumes a particular Ansatz for the subleading piece,
which does not fit well with our R-matrix. Somewhat relatedly, the Hamiltonians of our integrable models, as defined from the expansion of the transfer matrix, seems to have complicated expressions.}

Most of the integrable models in Figure \ref{Fig.zoo} gives rise to solutions to the standard YBE.

The exception is the case of 2d $\CN=(2,2)$ theory on $S^2$, where we obtain a generalization of the 
YBE called the ``cluster-enriched YBE'' in \cite{Yamazaki:2016wnu}.
Here we have a hybrid of the YBE with the transformation properties of the
``cluster $y$-variable'' \cite{FominZelevinsky4} in the cluster algebra \cite{FominZelevinsky1}.
Here the $S^2$ partition function is a non-trivial function on the set of FI parameters
associated to the vertices of the quiver diagram,
and the Seiberg-like duality holds only when we change the FI parameter,
whose transformation property coincides with the that in the cluster algebra \cite{Benini:2014mia}.

For 2d $\CN=(2,2)$ theory we obtain the standard YBE.
However, when we write the partition function \eqref{Z_localize}
there is a subtlety in the choice of the integration contour, and the correct prescription \cite{Benini:2013xpa}
is given by the Jeffrey--Kirwan residue \cite{MR1318878}. That we need to use such a residue prescription in the summation over `spins' in integrable models is a new feature of the integrable model.

\subsection{Super-Master Solution}

The case of 4d $\CN=1$ theory on $S^1\times S^3/\mathbb{Z}_r$ \cite{Benini:2011nc,Razamat:2013opa} is particularly interesting. 
The associated solution to the YBE \cite{Yamazaki:2013nra}, which we might call the ``super-master solution'' of the SSR (and hence YBE),
is one of the most general solutions to YBE ever known in the literature.

We can consider quiver gauge theories where the gauge group at each node is $\SU(N)$.
Then the model depends on two integers $N$ and $r$. The spins at each vertex takes values in
\be
(\mathbb{R}\times \mathbb{Z}_r)^{N-1} \;.
\ee
This arises from the holonomy of the gauge field, and the two factors $\mathbb{R}$ and $\mathbb{Z}_r$
correspond to the two generators of the fundamental group of $S^1\times S^3/\mathbb{Z}_r$:
\be
\pi_1(S^1\times S^3/\mathbb{Z}_r)= \mathbb{Z}\oplus \mathbb{Z}_r \;.
\ee

The partition function in this case is known as the lens index and was computed in \cite{Benini:2011nc} (see also \cite{Razamat:2013opa}).
In addition to the two integers $N$ and $r$ the partition function depends on
two parameters $p$ and $q$, which parametrizes the complex structure of 
the geometry $S^1\times S^3/\mathbb{Z}_r$ \cite{Kodaira}.

While integrability is a consequence of gauge
theory duality, integrability was recently proven mathematically
directly by \cite{Kels:2017toi}, by generalizing several earlier works
(\cite{Kels:2015bda} for $N=2, r>1$, \cite{RainsTransf} for $N\ge 2, r=1$ and $N=2, r=1$ \cite{SpiridonovBeta,MR2281166}). This is arguably the most impressive test of the 4d $\CN=1$ Seiberg duality in the literature.

One might imagine that 
the R-matrices for the super-master solution can be understood as 
intertwiners for some quantum-group like structure:\footnote{There is a possibility that the integer $r$ is a label for the 
representation of a single algebra $\mathcal{U}_{p,q}(\mathfrak{sl}_N)$. There are several indications, however, that this is not the case.}
\be
\mathcal{U}_{p,q;r}(\mathfrak{sl}_N) \;.
\label{Upqr}
\ee

For the special case of $r=1$ and $N=2$, it is known \cite{Zabrodin:2010qm} that this algebra is identified with the
Sklyanin algebra $\mathcal{U}_{p,q}(\mathfrak{sl}_2)$, an algebra defined by Sklyanin \cite{Sklyanin:1982tf} 
from the eight-vertex R-matrix \cite{Baxter:1972hz} via the so-called  RLL=LLR relation.\footnote{The L-operator for the case $r=1$ is identified as a BPS surface defect inside the 4d $\CN=1$ theory \cite{Maruyoshi:2016caf}.} It is natural to conjecture
that the algebra for $r=1, N>2$ is given by the $\mathfrak{sl}_N$-generalization of the 
Sklyanin algebra constructed by Cherednik \cite{CherednikGeneralized}. The algebra for the case of general $r>1$,
even for $N=2$, seems to be unknown. It is a fascinating question to 
identify the algebra \eqref{Upqr} for general $r$ and $N$.

\subsubsection{Root-of-Unity Limit}

Sine YBE is an equality, one can take appropriate limit of the parameters
and one could still hope that we have a solution to YBE.
The elliptic--trigonometric--rational degeneration we discussed above,
whose counterpart in quiver gauge theory is the dimensional reduction,
is a good example for this.

There is another particularly interesting limit 
worth mentioning. This is the root of unity limit. 

Such a limit has been discussed for example in the representation theory of quantum groups.
In our context, this limit is special since the Boltzmann weight diverges.

Let us first consider the case $N=2, r=1$ of the super-master solution. 
The Boltzmann weights
are given in terms of the elliptic gamma function (in the notation slightly different from \eqref{ell_gamma})
\be
\Phi(z;\p,\q) = \prod_{j,k=0}^{\infty} \frac{1-e^{2i z} \p^{2j+1} \q^{2k+1}}{1-e^{-2i z} \p^{2j+1} \q^{2k+1}} \;.
\ee
Let us consider the limit where $\q$ approaches a $2M$-th root of unity, while $\p$ stays finite:\footnote{Note that this integer $M$ is different from $N$,
which specifies the rank of the gauge groups.}
\be
\p=e^{i\pi \tau} \;, 
\quad
\q=e^{-\frac{\epsilon}{2 M^2}} \zeta \;,
\quad
\zeta^{2M}=1 \;.
\label{pq_limit}
\ee
Then $\Phi(z;\p,\q)$ diverges as \cite{Bazhanov:2010kz,Kels:2017vbc}\footnote{
See \cite{BazhanovReshetikhin,Garoufalidis:2014ifa,Ip:2014pva} for similar discussion for the quantum dilogarithm function.}
\be
\Phi(z;\p,\q)=\exp\left(
\frac{i}{\epsilon} 2M \int_0^z du \ln \overline{\vartheta}_3 (Nu| N\tau) 
\right) \times \textrm{(subleading finite piece)} \;,
\label{Phi_limit}
\ee
where  $\overline{\vartheta}_3(x|\tau)$ is the Jacobi theta function
\be
\overline{\vartheta}_3(x\,|\,\tau):= \prod_{n=1}^{\infty} (1+e^{2i x} e^{\pi i \tau (2n-1)}) (1+e^{-2 i x} e^{\pi i \tau (2n-1)}) \;.
\ee

This means that the Boltzmann weight diverges, which schematically reads
\be
Z \longrightarrow \sumint \prod_v d\sigma_v \,\, e^{\frac{1}{\epsilon} W^{(0)}(\sigma) + W^{(1)} (\sigma)+\mathcal{O}(\epsilon) }  \;.
\ee
In the limit $\epsilon\to 0$ we need to do the the saddle point analysis,
where the saddle point equation is given by
\be
\frac{\partial W^{(0)}(\sigma)}{\partial \sigma_v}=0 \;.
\ee
for each $v$.

Interestingly, for the case $N=2$ this saddle point equation in itself is identified with a discrete classical integrable equation
known as (Q4) in the list of Adler, Bobenko and Suris \cite{ABS}. In this paper the authors classified discrete integrable equations under several assumptions, and found that all their solutions arise from a degeneration of the most general equation (Q4).

The saddle point equation is also the vacuum equation for the 2d $\CN=(2,2)$ theory,
which is obtained from the 4d theory reduced on $T^2$. In the Gauge/Bethe correspondence of \cite{Nekrasov:2009uh}
this equation should be associated with the Bethe Ansatz equation for the integrable model.
One this note, however, this integrable model is different from ours. See \cite{Kels:2017vbc}
for further discussion on this point.\footnote{Note that the integrable models discussed in the work of Nekrasov and Shatashvili \cite{Nekrasov:2009uh}
and also of Costello \cite{Costello:2013zra} is the six-vertex model (Heisenberg XXZ spin chain) and its generalizations, which are themselves of different class from the chiral Potts models and their generalizations discussed in this lecture. However, there is a known connection between
chiral Potts model and the six-vertex model due to Bazhanov and Stroganov \cite{Bazhanov:1989nc},
and one expects that this will be the key for finding appropriate relations with the Gauge/YBE correspondence (see \cite{Yagi:2017hmj} for recent related 
discussion).}\footnote{See \cite{Sadri:2005gi} for yet another relation connection between quiver gauge theories and integrable spin chains.}

Now, once we have a solution to the saddle point equation we can then 
evaluate the subleading corrections, to obtain the YBE. By following this strategy
Bazhanov and Sergeev \cite{Bazhanov:2010kz} reproduced the Kashiwara--Miwa model \cite{Kashiwara:1986tu}
as well as the chiral Potts model \cite{vonGehlen:1984bi,AuYang:1987zc,Baxter:1987eq}.
In both cases the spins take values in $\mathbb{Z}_M$, where $M$ was introduced before in \eqref{pq_limit}.

This analysis was recently generalized to the case $r>1$ \cite{Kels:2017vbc} (still $N=2$). 
Since the expression for the Boltzmann weight is much more complicated than the $r=1$ case (in particular we already have $\mathbb{Z}_r$ discrete spins even before taking the limit)
we might expect to obtain a new solution of the YBE in the root of unity limit. We found a surprise in the root-of-unity limit of \eqref{pq_limit}:
after suitable change of variables, the $\mathbb{Z}_r$ spin, which was present before taking the limit, combines with the $\mathbb{Z}_M$ spin into a 
single $\mathbb{Z}_{Mr}$ spin. We moreover reproduced the same class of models as in the case of $r=1$ (namely Kashiwara--Miwa and chiral Potts models), with the only difference 
being $M$ is replaced by $Mr$. We hope that this observation would be of help in the search for the unknown algebraic structure $\mathcal{U}_{p,q;r}(\mathfrak{sl}_N)$ discussed in \eqref{Upqr}.

It is rather interesting that the chiral Potts model arises from the root-of-unity limit of the partition function of supersymmetric quiver gauge theories.
The chiral Potts models is special in that the spectral parameters take values in the higher-genus spectral curve.
Another peculiar feature of the model is that it does not have the rapidity difference property \eqref{rapidity_difference}.
Partly due to these peculiarities the model has long defied understanding from quantum field theory. For example, in the 1989 paper \cite{Witten:1989wf} Witten wrote,
in the paper where he discussed his approach to integrable models from three-dimensional Chern-Simons theory:

\begin{quotation}
There are several obvious areas for further investigation. \dots Another question, which may or may not be related, is to understand the spin models formulated only rather recently in [24] in which the spectral parameter is not an abelian variable (as in previous constructions),
but is a point on a Riemann surface of genus greater than one. 
\end{quotation}

In this quote, reference [24] points to the literature on chiral Potts models. One might be able to say that the Gauge/YBE correspondence finally provided an answer
to this question. Interestingly, this paragraph also raises a few more open questions, which reads (again quoting from \cite{Witten:1989wf})

\begin{quotation}
In terms of statistical mechanics, one compelling question is to understand the origin of the spectral parameter
(and the elliptic modulus) in IRF and vertex models; this is essential for explaining the origin of integrability. \dots If possible, one would also like to understand vertex models and quantum groups directly at physical values of $q$, without the less than appealing analytic continuation that is used in this paper. 
\end{quotation}

Both these questions are answered in our framework; the origin of the spectral parameter is the R-charge in supersymmetric gauge theories (recall section \ref{sec.spectral}),\footnote{One should add that the Costello's approach \cite{Costello:2013zra,Costello:2013sla} also answers this question, in the setup much closer than us to the original 3d Chern-Simons setup of \cite{Witten:1989wf}.}
and the value of the parameter $\q$ (and one more parameter $\p$) are parameters for the complex structure moduli and hence are continuous parameters from the outset in our context .

\section{Conclusion}

In this lecture we reviewed the subject of the Gauge/YBE correspondence.
For the convenience of the readers we can summarize the 
dictionary between integrable models and quiver gauge theories in Table \ref{tab.summary}.

\begin{table}[htbp]
\caption{Dictionary for the Gauge/YBE correspondence.}
\begin{center}
{\renewcommand\arraystretch{1.2}
\begin{tabular}{c|c}
integrable model & quiver gauge theory \\
\hline
spin lattice & quiver diagram \\
chiral & chiral \\
rapidity line & zig-zag path \\
spectral parameter & R-charge \\
statistical partition function & supersymmetric partition function \\
temperature-like parameters & quantum parameters (such as $\p, \q$) \\
spin variables & gauge holonomies along non-contractible cycles  \\
self-interaction & gauge multiplet \\
nearest-neighbor interaction & bifundamental matter multiplet  \\
star--star relation & Seiberg(-like) duality \\
R-matrix & theory $\mathcal{T}[R]$ (Figure \ref{Fig8_1}) \\
composition of R-matrices & gluing $\mathcal{T}[R]$'s by gauging\\
Yang-Baxter equation & Yang-Baxter duality (Figure \ref{Fig8_4})\\
degeneration & dimensional reduction
\end{tabular}
}
\end{center}
\label{tab.summary}
\end{table}

I began this lecture by asking the question: why integrable models exist?

The Gauge/YBE correspondence provides an elegant answer for this question:
integrable models (solutions to Yang-Baxter equation) exist because
there are underlying gauge theory dualities, namely the Yang-Baxter dualities.

One can still try to go deeper, and ask the question of why gauge theory dualities exist.
One possible answer is that the gauge `symmetry' is really not a symmetry but rather a
redundancy of the description, introduced to make manifest many of the fundamental principles of quantum field theory,
such as locality and unitarity. This means that integrability is ultimately tied with 
locality and unitarity. Such an understanding might contain deep insight into the true origin of integrable models.

One novel aspect of the Gauge/YBE correspondence is that
the integrability resides not in each individual quiver gauge theory, but in the ``theory space'' spanned by a {\it class} of quiver gauge theories
glued together by gauging---the conserved charges of integrable models maps one theory to another.
This might be the indication that there are many unknown structures yet to be discovered inside the theory space,
see \cite{YamazakiBook} for exposition of the relevant idea and \cite{Yamazaki:2013xva,Benini:2014mia,Terashima:2013fg} for some related attempts.

In conclusion, I hope that in these lectures I have convinced the reader of the 
richness of the subject of the Gauge/YBE correspondence. The correspondence
has lead us to new physics and new mathematics in the past,
and I am convinced that it will continue to do so in the years to come.

\section*{Acknowledgements}

The author would like to take this opportunity to thank 
my collaborators in various stages of this research, and to the organizers of the String-Math 2018 conference for 
providing a stimulating environment.
The underlying material for this lecture has been presented in several workshops,\footnote{For example,
``Workshop on Geometric Correspondences of Gauge Theories'', ICTP, Sep.\ 2013;
``Integrability in Gauge and String Theory (IGST 2014)'', DESY, Jul.\ 2014;
``IGST 2015'', King's College, Jul.\ 2015;
``Baxter 2015: Exactly Solved Models and Beyond'', Palm Cove, Australia, Jul.\ 2015.}
and he would like to thank the audience for feedback.
This research was supported in part by World Premier International Research Center Initiative (WPI Initiative), MEXT, Japan.,
and by JSPS Grant No.\ 17KK0087.

\bibliographystyle{nb}
\bibliography{YBE_review}


\end{CJK*}


\end{document}